\documentclass[12pt]{iopart}
\usepackage{graphicx}

\usepackage{csquotes}

\expandafter\let\csname equation*\endcsname\relax

\expandafter\let\csname endequation*\endcsname\relax

\usepackage{amsmath}
\usepackage{lipsum}
\usepackage{float}
\usepackage{xcolor}
\usepackage{multirow}
\usepackage{siunitx}

\usepackage[backend=biber,style=authortitle,bibstyle=numeric, citestyle=phys,doi=false,url=false,isbn=false, maxnames = 20]{biblatex}

\DeclareMathOperator{\SBR}{SBR}
\DeclareMathOperator{\SNR}{SNR}

\begin{document}

\title[Spectral dose imaging]{Technical note: On the use of polychromatic cameras for high spatial resolution spectral dose measurements}

\author{E Cloutier$^{1,2}$, L Beaulieu$^{1,2}$,
L Archambault$^{1,2}$}
\address{$^1$ Service de physique médicale et Axe Oncologie du Centre de recherche, CHU de Québec-Université Laval, Canada}
\address{$^2$ Département de physique, de génie physique et d'optique, et Centre de recherche sur le cancer, Université Laval, Québec, Canada}

\begin{abstract}
\\
    Despite the demonstrated benefits of hyperspectral formalism for stem effect corrections in the context of fiber dose measurements, this approach has not been yet translated into volumetric measurements where cameras are typically used for their distinguishing spatial resolution. This work investigates demosaicing algorithms for polychromatic cameras based spectral imaging. The scintillation and Cherenkov signals produced in a radioluminescent phantom are imaged by a polychromatic camera and isolated using the spectral formalism. To do so, five demosaicing algorithms are investigated from calibration to measurements: a clustering method and four interpolation algorithms. The resulting accuracy of scintillation and Cherenkov images is evaluated with measurements of the differences (mean $\pm$ standard deviation) between the obtained and expected signals from profiles drawn across a scintillation spot. Signal-to-noise ratio and signal-to-background ratio are further measured and compared in the resulting scintillation images. Clustering, OpenCV, bilinear, Malvar and Menon demosaicing algorithms respectively yielded differences of $3\pm6\%$, $0.1\pm0.5\%$, $0.5\pm0.5\%$,  $1\pm3\%$ and $1\pm4\%$ in the resulting scintillation images. For the Cherenkov images, all algorithms provided differences below 1\%. All methods enabled measurements over the detectability ($\SBR>2$) and sensitivity ($\SNR>5$) thresholds with the bilinear algorithm providing the best SNR value. Hence, radioluminescent signals can accurately be isolated using a single polychromatic camera. Moreover, demosaicing using a bilinear kernel provided the best results and enabled stem effect subtraction while preserving the full spatial resolution of the camera.  
    
\end{abstract}

\maketitle

\section{Introduction}

Scintillation detectors have evolved from single point measurement devices to volumetric detectors enabling 2D and even 3D measurements \cite{beaulieu_review_2016}. The use of fluors, in the production of scintillation detectors, whether incorporated into scintillating fibers, liquid scintillators or solid plastic bulk, is motivated by the resulting scintillators water equivalence over a wide range of therapeutic energies, which enables correction-free measurements \cite{beddar_water-equivalent_1992, beddar_water-equivalent_1992-1}. In addition, these scintillators can perform real-time measurements with a high spatial resolution. This makes plastic scintillation detectors (PSD) attractive for small fields or highly modulated beam measurements \cite{Xue_2017_small_field}. 

In the last few years, many photodetectors were proposed to measure the scintillation signal, each having specific strengths and weaknesses. Selecting the optimal photodetector implies making trade-offs between spatial resolution, spectral resolution and dose rates operating ranges \cite{boivin_systematic_2015}. While the operating range determines the accuracy of a measurement, spatial and spectral resolutions increase its reach: a high spatial resolution will enable dose sampling from many points in the volume while spectral resolution is advantageous for stem-effect removal and measurements with multi-point probes. 

A spectral mathematical formalism has been specifically developed to isolate the signals of interest in PSD dose measurements \cite{archambault_mathematical_2012}. The so-called hyper-spectral/multi-spectral formalism was successfully applied to multi-point PSD  measurements as well as stem-effect and temperature-dependence removal \cite{therriault-proulx_development_2012, linares_rosales_optimization_2019, therriault-proulx_method_2015}. In that context, a photomultiplier tubes (PMTs) assembly and spectrometers were used to acquire the spectral information. These detectors have a high sensitivity, sensibility and spectral resolution, but provide limited spatial resolution. Cameras, for their part, have enabled high spatial resolution 2D and 3D dose measurements \cite{goulet_novel_2014, kirov_three-dimensional_2005, alexander_1_nodate}. Taking advantage of color channels, together with the spectral formalism, polychromatic cameras have the potential of combining many photodetector qualities into one device. 

Polychromatic cameras typically acquire spectral information using three color channels (red, blue, green) integrated as mosaics overlaid on the sensor. Hence, the spectral information is not acquired at each pixel and is instead sampled to form incomplete color matrices. The purpose of this study is to investigate the combined use of demosaicing algorithms with spectral formalism for spectral dose imaging. 

\section{Theory and Methods}
\subsection{Spectral imaging}
Spectral approaches are based on the idea that the signal results from a linear superposition of spectra \cite{archambault_mathematical_2012}. In that case, a measurement can be described as:
\begin{align}
\boldsymbol{M} = \boldsymbol{R}\cdot \boldsymbol{D} 
\label{eq:multi}
\end{align}
where $\boldsymbol{M}$ is the measurement matrix, $\boldsymbol{R}$ is the detector response and $\boldsymbol{D}$ is the dose. The detector response is the matrix containing the individual spectra from each radio-luminescent element as measured by the optical system and obtained from calibration. Then, from equation~\ref{eq:multi}, the dose can be extracted from a measurement using the prior detector response:
\begin{align}
    \boldsymbol{D} = \boldsymbol{R}^+\boldsymbol{M}.
\end{align}
where $\boldsymbol{R}^+$ is the Moore-Penrose pseudo inverse (i.e. $(\boldsymbol{R}^T \boldsymbol{R})^{-1}\boldsymbol{R}^T$). For the demonstrated case of a scintillating probe, the spectra are either read by a series of color sensitive detectors (discrete spectra) \cite{linares_rosales_optimization_2019} or a spectrometer (near-continuous spectra) \cite{therriault-proulx_development_2012, Jean_2021}. In that context, each channel, or spectral band, measures the spectral information coming from the same sensitive volume. Spectral acquisition using a polychromatic camera complexifies the approach as the spectral information is not sampled at each pixel, but rather forms incomplete color matrices.

\begin{figure}[ht]
    \centering
    \includegraphics[width = 0.6\textwidth]{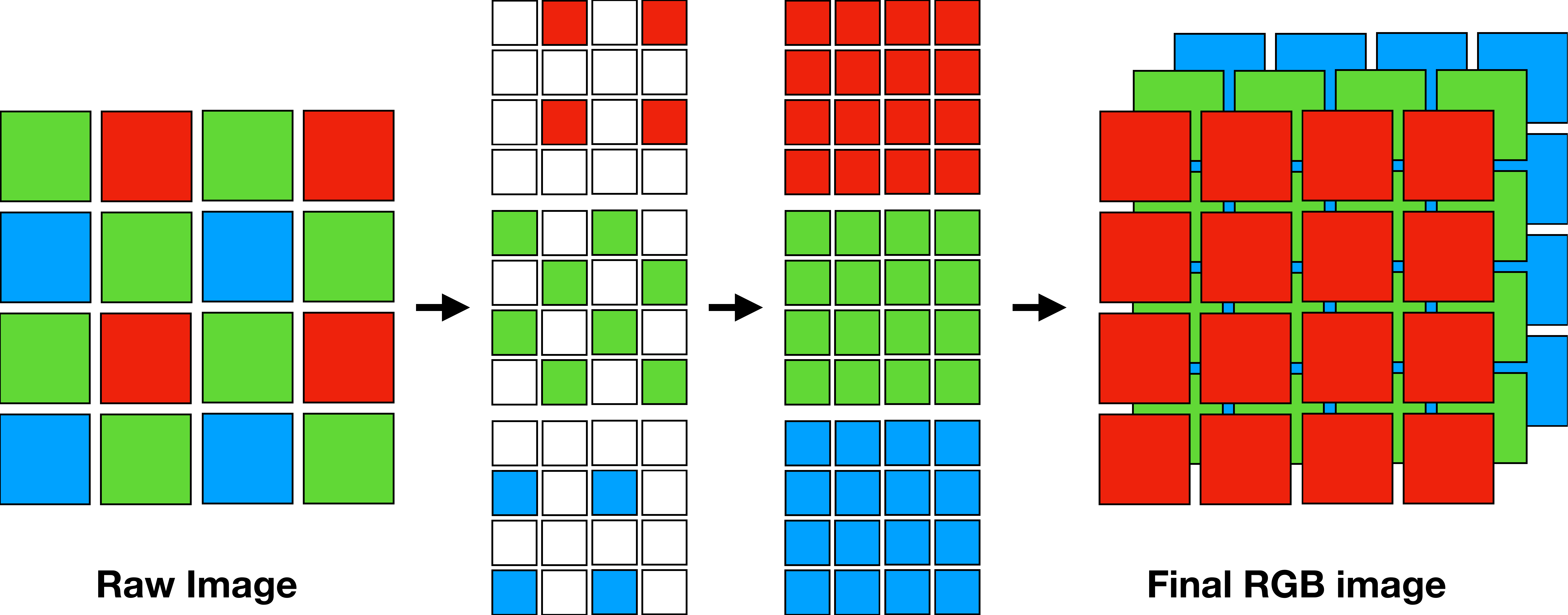}
    \caption{The left panel presents a 4x4 pixel portion of a Bayer camera filter. From left to right is the workflow scheme of demosaicing where the objective is to retrieve complete color matrices. }
    \label{fig:demosaicing}
\end{figure}

In this work, we used a polychromatic camera with a Bayer filter (Alta U2000, Andor Technology, Belfast, United Kingdom). Each block of 4 pixels contains two green, one red and one blue sensitive pixels as presented on figure~\ref{fig:demosaicing}.
To retrieve the complete spectral information, one has to work with incomplete color mosaics for each color. Processing these matrices into a full-color image is called demosaicing. Different demosaicing algorithms were tested, each of which can be classified in one of two categories: clustering and interpolation methods. Each demosaicing algorithm was tested from calibration to dose measurements. 

\subsubsection{Clustering Method}

A first approach is to cluster pixels into groups of $2\times2$ pixel comprising one blue (B), one red (R) and two green (G) pixels from the Bayer pattern of the sensor. Green pixel values are averaged. Spectra consisted of the mean R/G, G/G and B/G ratios taken over a region of interest. The strength of this method is that it does not rely on interpolation that may bias the results. However, it assumes that the same signal is seen by the individual pixels, which might not hold true in regions of high dose gradients. Furthermore, it reduces the camera's spatial resolution by a factor of two.

\subsubsection{Interpolation Methods}

Another approach is to estimate the missing color contributions to a pixel using interpolation from the neighboring pixels. For example, a blue pixel will be interpolated from surrounding green and red values. In that case, green values can be expected to be better estimated as it is sampled twice more often than the other colors in a Bayer mosaics. Four interpolation-based demosaicing algorithms were tested : the OpenCV demosaicing algorithm \cite{opencv_library}, a bilinear kernel algorithm, the demosaicing methods published by Malvar and al. \cite{Malvar_2004} and that published by Menon and al. \cite{Menon_2007}. The OpenCV and bilinear algorithms both use the following convolution kernels for green $\mathcal{K}_G$, blue and red $\mathcal{K}_{R/B}$, pixels interpolation:
\begin{align}
    & \mathcal{K}_G = \frac{1}{4} \cdot \begin{bmatrix}
    0 & 1_G & 0 \\
    1_G & 4 & 1_G \\
    0 & 1_G & 0\end{bmatrix}  && \mathcal{K}_{R/B} = \frac{1}{4} \cdot \begin{bmatrix}
    1_{R/B} & 2_{R/B} & 1_{R/B} \\
    1_{R/B} & 4 & 2_{R/B} \\
    1_{R/B} & 2_{R/B} & 1_{R/B} 
    \end{bmatrix}
\end{align}
The  $\mathcal{K}_R$, $\mathcal{K}_G$ and  $\mathcal{K}_B$ kernels are respectively applied to the incomplete red, green and blue matrices. The approaches developed by Malvar and Menon, which were developed for photography, further aim at using inter-color correlation, i.e. information from other surrounding colors, to estimate each contribution. For example, the algorithm published by Malvar et al. defines convolution kernels for estimating the green value at a blue or red pixel $\mathcal{K}_{G\rightarrow R/B}$, the red or blue value at a green pixel in a red or blue line  $\mathcal{K}_{R/B\rightarrow G_l}$, the red or blue value at a green pixel in a red or blue column $\mathcal{K}_{R/B\rightarrow G_c}$ and the red or blue value at a blue or red pixel $\mathcal{K}_{R/B\rightarrow B/R}$ \cite{Malvar_2004}:
\begin{align}
     \mathcal{K}_{G\rightarrow R/B} &= \frac{1}{8} \cdot \begin{bmatrix}
    0 & 0 & -1_{R/B} & 0 & 0 \\
    0 & 0 & 2_G & 0 & 0 \\
    -1_{R/B} & 2_G & 4_{R/B} & 2_G & -1_{R/B} \\
    0 & 0 & 2_G & 0 & 0 \\
    0 & 0 & -1_{R/B} & 0 & 0
    \end{bmatrix} \\
      \mathcal{K}_{R/B\rightarrow G_l} &= \frac{1}{8} \cdot \begin{bmatrix}
    0 & 0 & 0.5_{G} & 0 & 0 \\
    0 & -1_{G} & 0 & -1_{G} & 0 \\
    -1_{G} & 4_{R/B} & 5_G & 4_{R/B} & -1_{G} \\
    0 & -1_{G} & 0 & -1_{G} & 0 \\
    0 & 0 & 0.5_{G} & 0 & 0
    \end{bmatrix} \\
    \mathcal{K}_{R/B\rightarrow G_c} &= (\mathcal{K}_{R/B\rightarrow G_l})^t \\
         \mathcal{K}_{R/B\rightarrow B/R} &= \frac{1}{8} \cdot \begin{bmatrix}
    0 & 0 & -1.5_{B/R} & 0 & 0 \\
    0 & 2_{R/B} & 0 & 2_{R/B} & 0 \\
    -1.5_{B/R} & 0 & 6_{B/R} & 0 & -1.5_{B/R} \\
    0 & 2_{R/B} & 0 & 2_{R/B} & 0 \\
    0 & 0 & -1.5_{B/R} & 0 & 0
    \end{bmatrix} 
\end{align}
By introducing these kernels, the Malvar algorithm estimates the missing pixel values using gradient correction interpolations. Finally, the demosaicing method proposed by Menon et al. further uses gradients estimated by green, red and blue values to predict all colors, in a complex 4-step algorithm~\cite{Menon_2007}. First, two green matrices are constructed using directional, horizontal or vertical, interpolation. The resulting $G^H$ and $G^V$ matrices are then used to predict the gradient direction and to guide the red and blue components interpolation. Finally, a refining step corrects interpolation artifacts in all three channels caused by the edge direction selection. For more details, the reader is encouraged to refer to the original manuscripts~\cite{Malvar_2004,Menon_2007}.

\subsection{Calibration and Dose Measurements}

\begin{figure}[ht]
    \centering
    \includegraphics[width = 0.6\textwidth]{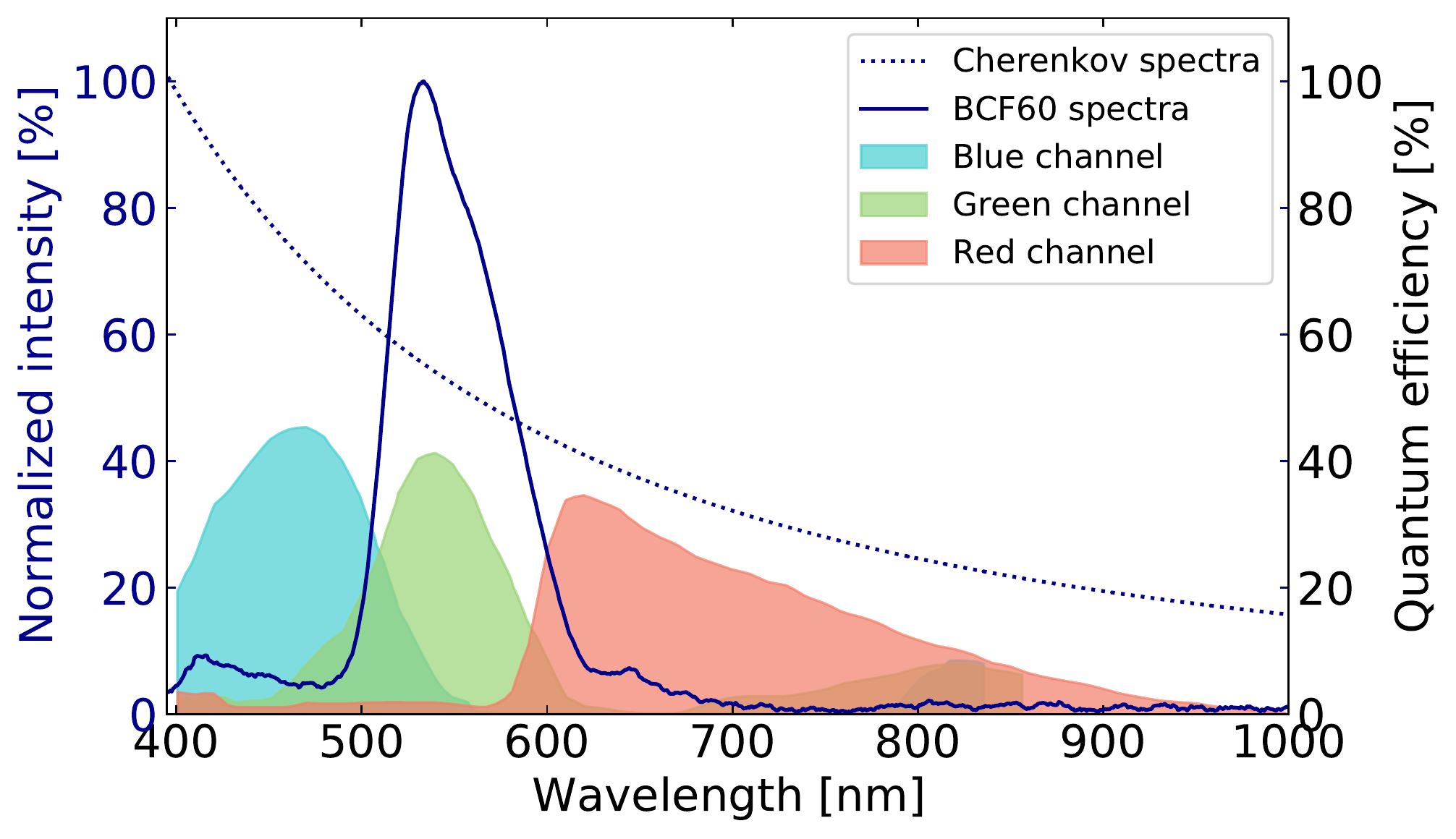}
    \caption{Scintillation and Cherenkov radiation normalized spectra (left axis) along with the colored channel quantum efficiency of the Alta U2000 CCD camera (right axis).}
    \label{fig:camera_filter}
\end{figure}

Spectral imaging using demosaicing methods were tested using a phantom that consists of a circular disk of transparent plastic containing an array of 19 green plastic scintillators (BCF-60; Saint-Gobain Crystals, Hiram, OH, USA)~\cite{cloutier_deformable_2021_1, cloutier_deformable_2021_2}. The phantom is subject to the emission of two signals that need to be distinguished: the scintillation signal produced from the fibers and the Cherenkov stem signal produced in the bulk. Figure~\ref{fig:camera_filter} presents both radioluminescent spectra of interest together with the camera channels quantum efficiency. 
The phantom bulk is made of a urethane compound (ClearFlex 30; Smooth-On, Macongie, USA) cast in a 6\,cm diameter and 1.2\,cm thick circular mold. The scintillating fibers are 1.2\,cm long and have a 1\,mm diameter. The scintillating fibers were embedded in the phantom so that they are 1.5\,cm equidistant from each other. For calibration, scintillation spectra were obtained by irradiating the phantom under a 120 kVp orthovoltage photon beam. This energy is below the Cherenkov threshold and thus ensures that only scintillation is acquired, neglecting fluorescence. To measure the Cherenkov emission spectra, an additional disk was manufactured, having the same dimensions as the first one, except that no scintillating fibers were inserted. The Cherenkov spectrum was acquired by placing that additional phantom at the isocenter of a 6 MV photon beam. For measurements, the scintillating phantom was aligned at the isocenter of a $6\times2 \text{ cm}^2$, 6 MV beam. Considering the RBG channels of the camera together with the two radioluminescent sources, equation~\ref{eq:multi} becomes:
\begin{align}
    \begin{bmatrix}
    M_R \\
    M_G \\
    M_B 
    \end{bmatrix} = \begin{bmatrix}
    R_{R, BCF60} & R_{R, Cherenkov} \\
    R_{G, BCF60} & R_{G, Cherenkov} \\
    R_{B, BCF60} & R_{B, Cherenkov}
    \end{bmatrix} \cdot \begin{bmatrix}
    D_{BCF60} \\
    D_{Cherenkov}
    \end{bmatrix}.
\end{align}

For each measurement, the camera was positionned on the treatment couch, 50 cm from the isocenter. A 12\,mm focal length objective lens was coupled to the camera. Five frames of 1\,s were acquired. Background frames, i.e. images in absence of radiation, were subtracted from the signal images and median temporal filter, over five acquisitions, to further correct the remaining transient noise~\cite{archambault_transient_2008}. For each demosaicing algorithm tested, a signal-to-noise and signal-to-background analysis was performed to quantify the sensitivity and detectability of the resulting images of scintillation. These were defined as:
\begin{align}
   \SBR = \frac{\mu_{spot}}{\sigma_{bg}}, \qquad \SNR_{avg} = \frac{\mu_{s}}{\sigma_{s}}, \qquad \SNR_{spot}= \sqrt{n}\SNR_{avg}.
\end{align}
$\SNR_{avg}$ is the SNR where pixels from each scintillating fiber is treated individually whereas $\SNR_{spot}$~\cite{lacroix_design_2009} is that when pixels forming a spot are treated as a group. Finally, the crest factor was further computed to quantify the detectability of the scintillation signal over the stem effect. The crest factor is defined as the peak to average ratio (PAR) where the peak is determined by the scintillation and the average is the remaining stem effect. Higher PAR are expected for methods that accurately removed the stem contribution on scintillation images.
\begin{figure}[ht]
    \centering
    \includegraphics[width = 0.99\textwidth]{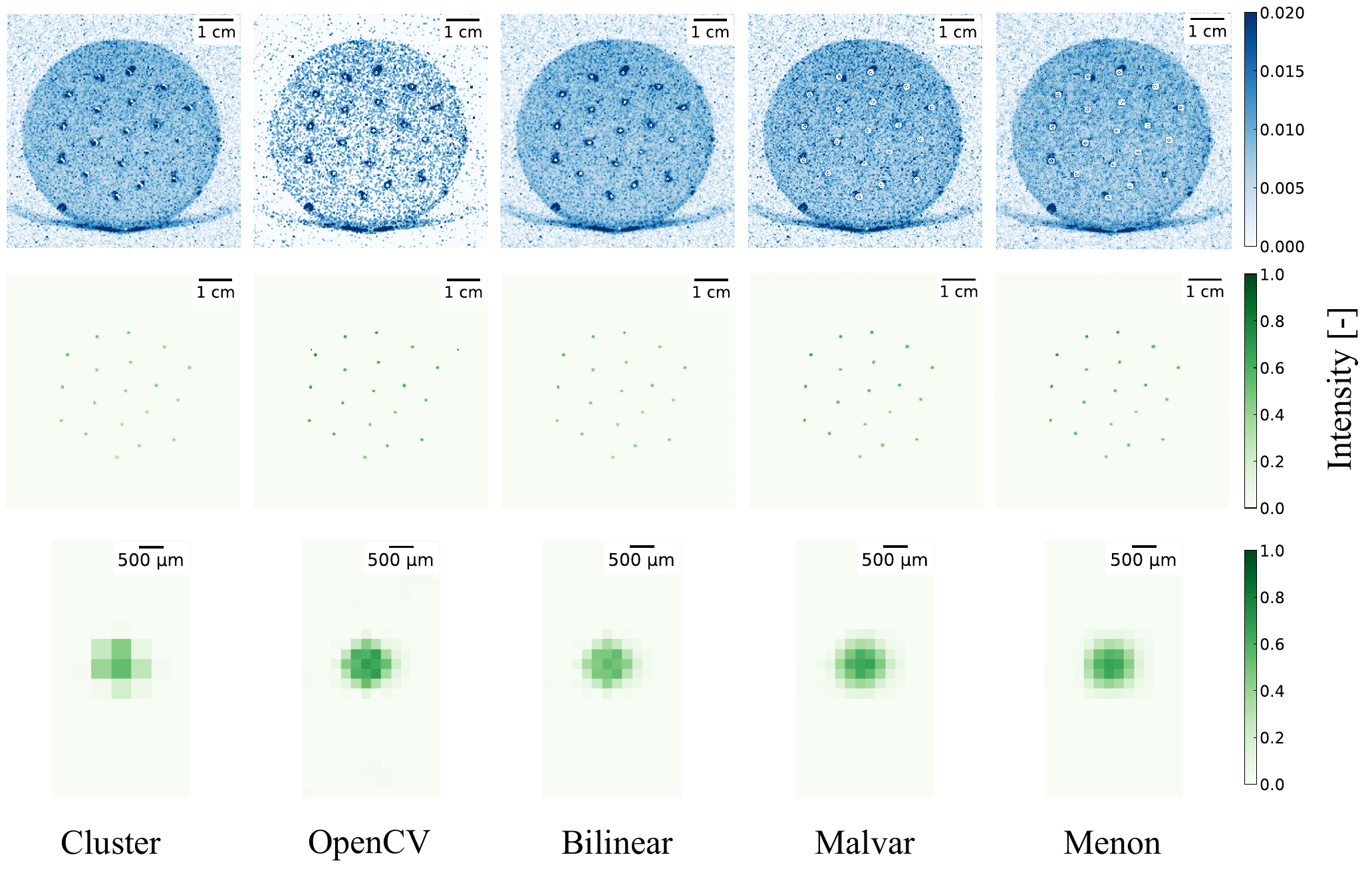}
    \caption{Resulting dose images after spectral deconvolution using the different demosaicing algorithms. Top panel present the Cherenkov signal, middle panel the scintillation signal whereas bottom panel is a zoom on the central scintillation spot.}
    \label{fig:pixel_demosaicing}
\end{figure}
\section{Results and Discussion}

Figure~\ref{fig:pixel_demosaicing} presents Cherenkov and scintillation dose images that were obtained using the different demosaicing algorithms. In all cases, the spectral method isolated the Cherenkov signal, mainly produced in the bulk, from the scintillation signal generated from the 19 scintillating fibers. Therefore, scintillating fibers locations are visible as lower intensity regions on the Cherenkov images whereas the Cherenkov signal is totally removed from the scintillating images. Looking at the scintillation spots obtained with the clustering method, the figure highlights the spatial resolution reduction imposed by this method. 

\begin{figure}[ht]
    \centering
    \includegraphics[width = 0.9\textwidth]{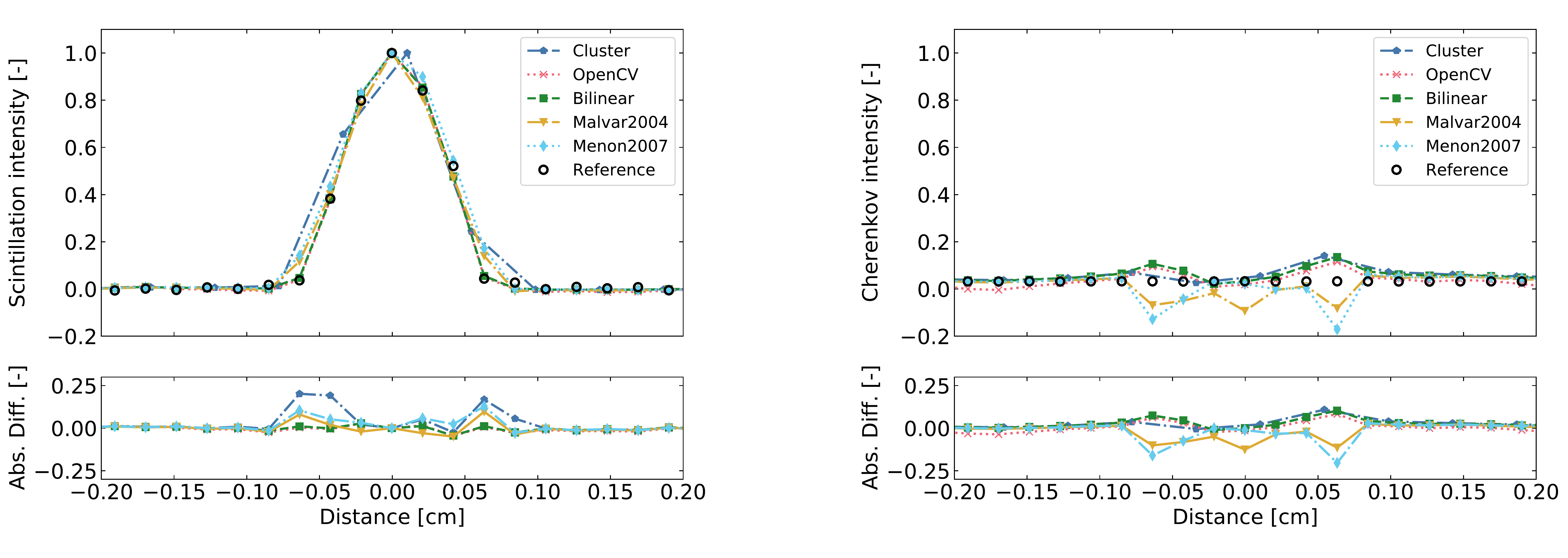}
    \caption{Profiles drawn across a scintillation spot in both scintillation (left) and Cherenkov (left) images obtained with the different demosaicing methods.}
    \label{fig:spot_analysis}
\end{figure}

For the dose accuracy, figure~\ref{fig:spot_analysis} presents profiles drawn across a scintillating spot compared to the expected signal for both scintillation and Cherenkov images. For the scintillation images, differences up to 20\% are obtained using the clustering algorithm, which are attributed to volume averaging being more present. Differences of (mean $\pm$ standard deviation) $3\pm6\%$ are obtained using the clustering algorithm on the scintillation images. The bilinear and OpenCV algorithms presented differences of  respectively $0.5\pm0.5\%$ and $0.1\pm0.5\%$. Similar results were expected for these algorithms since demosaicing is performed along the same convolution kernels. However, it was found that the OpenCV routine rounds its output values, which slightly biases the calibration spectra. This explains the differences between the bilinear and OpenCV results. Malvar and Menon algorithms presents differences of $1\pm3\%$ and $1\pm4\%$ respectively, as the spot width seems to be overestimated. For the Cherenkov images, comparison is trickier since a reference is hard to establish at the scintillating fibers location, not knowing the true Cherenkov contribution in the scintillating volume. Hence, the reference Cherenkov signal is the one obtained from the bulk without fibers. Still, outside of the fibers region of interest, this contribution is well estimated by most algorithms, i.e., differences remained below $\pm1\%$. Overall, similar performances are obtained within the same family of algorithms, namely the clustering, bilinear, and inter-channel correlation (Malvar and Menon) based algorithms.

\begin{figure}[ht]
    \centering
    \includegraphics[width = 0.9\textwidth]{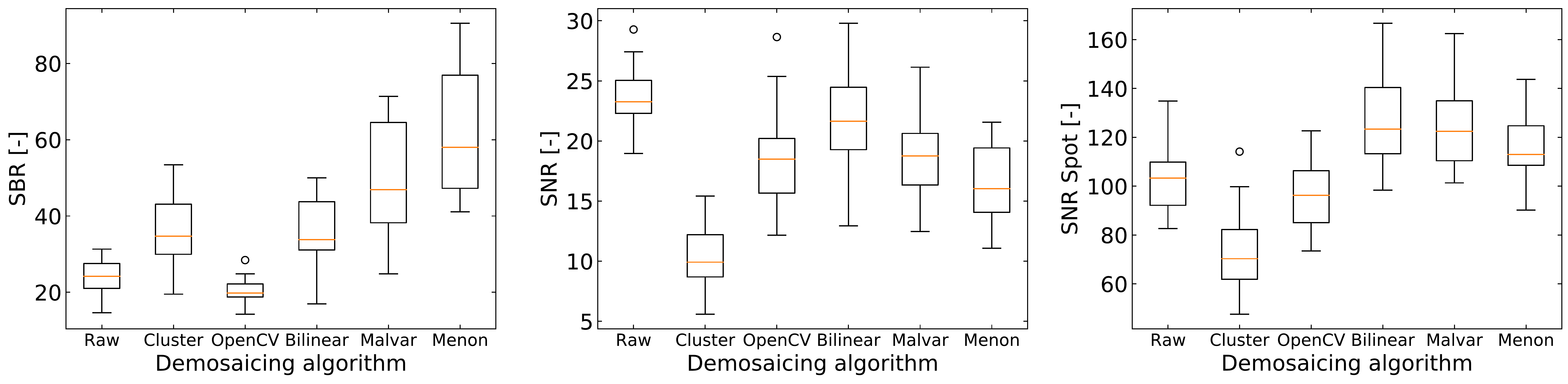}
    \caption{SBR and SNR analysis conducted on the scintillation images generated with the different demosaicing algorithms. Raw measurements correspond to the analysis of the images not corrected for Cherenkov signals. The spread results from the analysis on the 19 scintillating fibers. }
    \label{fig:snr_sbr}
\end{figure}

\begin{figure}[ht]
    \centering
    \includegraphics[width = 0.35\textwidth]{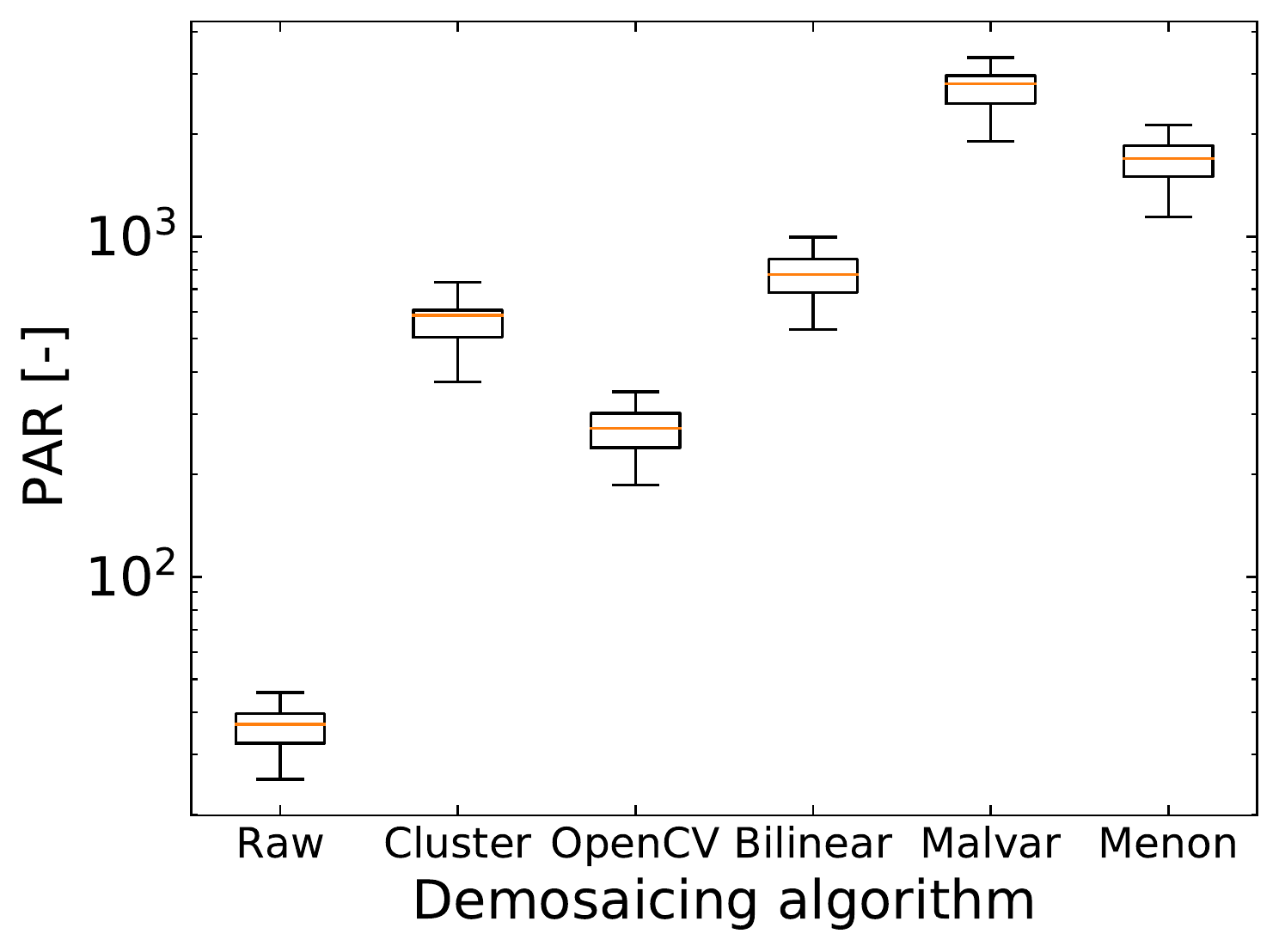}
    \caption{Peak to average ratio (PAR) on the images generated with the five demosaicing algorithms in comparison to the raw images. For each algorithm, analysis was conducted over the 19 scintillating spots.}
    \label{fig:par}
\end{figure}

The resulting sensitivity and detectability of scintillation images is presented in figure~\ref{fig:snr_sbr} where SNR and SBR analysis are conducted over the 19 scintillating fibers. Overall, the Cherenkov subtraction from the images resulted in higher SBR, or detectability, in the scintillation images, except for the OpenCV method. This reinforces the benefits of spectral Cherenkov correction. Moreover higher SBR are obtained using Malvar and Menon algorithms which is attributed to a higher portion of the signal falling into the scintillation dose channels. As for the SNR analysis, bilinear demosaicing resulted in the higher performances, while the clustering method degraded the SNR from raw images. Still, for all methods, SNR and SBR remained higher than the detectability ($\SBR>$2) and sensitivity ($\SNR>$5) thresholds. SBR analysis was conducted using the standard deviation of the noise $\sigma$, which corresponds to a region of interest without any source of light. On the scintillation images, similar $\sigma$ were found in the bulk region where Cherenkov was removed. 
As for the scintillation detectability over stem effect, figure~\ref{fig:par} presents the PAR of the raw image in comparison to the scintillation images obtained with the different demosaicing algorithms. Stem effect removal through the hyperspectral formalism increased the detectability of the scintillating fibers by an order of magnitude in all cases when compared to raw images not corrected for Cherenkov emission.  

Overall, the demosaicing techniques provided Cherenkov and scintillation signal separation in good agreement with the expected dose. The bilinear algorithm presented the best performance with regards to the dose spatial accuracy, detectability and sensitivity. The bilinear kernel enabled scintillation and Cherenkov signal separation, within $1\%$ agreement, while preserving the full spatial resolution of the camera. Furthermore, SBR and spot SNR were enhanced following the bilinear deconvolution compared with the raw images.  

Previous studies using volumetric scintillators typically treated Cherenkov removal by subtraction of the set-up's contribution~\cite{Frelin_dosemap_2008,goulet_novel_2014, rilling_tomographicbased_2020, delage_characterization_2018}. In particular, Frelin et al. proposed a Cherenkov discrimination technique based on spatial modulation imposed by a semi-transparent chess-board pattern~\cite{Frelin_dosemap_2008, dosimap_2009}. However, these techniques cannot take into account Cherenkov produced directly within the scintillating material. Moreover, these approach are more difficult to apply for phantoms that deform during measurements. Deformable scintillation-based dosimeters~\cite{cloutier_deformable_2021_1, cloutier_deformable_2021_2} could especially benefit from robust 2D and 3D Cherenkov signal removal. Use of the spectral formalism proposed in this study enabled the separation of the scintillation signal from the Cherenkov stem signal while preserving spatial information. 

Cherenkov filtering using cameras was initiated for scintillation probe measurements~\cite{Frelin_2005, lacroix_2008_clinical,guillot_spectral_2011,guillot_performance_2013}. In particular, the method presented by Lacroix et al. first proposed spectral measurements using a polychromatic camera, but only took advantage of the blue and green pixels, treated in clusters~\cite{lacroix_2008_clinical}. Approaches presented by Frelin et al. and then Guillot et al. used two monochromatic cameras in front of which a dichroic filter divided the raw signal into two contributions reaching each of the cameras~\cite{Frelin_2005,guillot_spectral_2011, guillot_performance_2013}. Frelin et al. also presented a setup using a single monochromatic camera where measurements were reproduced through different spectral bands~\cite{Frelin_dosemap_2008}. Recently, a Cherenkov spectral analysis was conducted along the same lines, whereas the spectral information was simultaneously acquired from three cameras sensitive to different spectral bands~\cite{alexander_color_2021}. Hence, for these previous techniques, no demosaicing techniques were necessary. In our case, the combined use of Bayer filter cameras and demosaicing algorithms provides the advantage of a simplified measuring set-up where a single camera is necessary. The approach here is also different since the method takes advantage of all three colored channels of Bayer cameras and relies on individual calibration of the different spectra. This is expected to make the method more robust and dose measurements more accurate.

\section{Conclusion}
Demosaicing algorithms were investigated in the context of spectral imaging using polychromatic cameras. Image demosaicing using a bilinear kernel presented the best performance, resulting in accurate spatial dose measurements and increased detectability and sensitivity in comparison to the raw images. Together with a polychromatic CCD camera, the Cherenkov signal was accurately isolated from the scintillation signal using the spectral formalism. Given these results, the combined use of demosaicing algorithms and spectral formalism makes polychromatic cameras a promising tool for improved volumetric dose measurements.

\section{Acknowledgement} 

The authors thank Ghyslain Leclerc for the English revision of the manuscript. This work was financed by the Natural Sciences and Engineering Research Council of Canada (NSERC) Discovery grants \#2019-05038 and \#2018-04055. Emily Cloutier acknowledges support by the Fonds de Recherche du Quebec – Nature et Technologies (FRQNT). 


\clearpage

\begin{small}
\printbibliography[heading=bibintoc]

@article{Frelin_2005,
author = {Frelin, A-M. and Fontbonne, J-M. and Ban, G. and Colin, J. and Labalme, M. and Batalla, A. and Isambert, A. and Vela, A. and Leroux, T.},
title = {Spectral discrimination of Čerenkov radiation in scintillating dosimeters},
journal = {Medical Physics},
volume = {32},
number = {9},
pages = {3000-3006},
doi = {https://doi.org/10.1118/1.2008487},
url = {https://aapm.onlinelibrary.wiley.com/doi/abs/10.1118/1.2008487},
eprint = {https://aapm.onlinelibrary.wiley.com/doi/pdf/10.1118/1.2008487},
year = {2005}
}

@article{dosimap_2009,
author = {Collomb-Patton, V. and Boher, P. and Leroux, T. and Fontbonne, J.-M. and Vela, A. and Batalla, A.},
title = {The DOSIMAP, a high spatial resolution tissue equivalent 2D dosimeter for LINAC QA and IMRT verification},
journal = {Medical Physics},
volume = {36},
number = {2},
pages = {317-328},
doi = {https://doi.org/10.1118/1.3013703},
url = {https://aapm.onlinelibrary.wiley.com/doi/abs/10.1118/1.3013703},
eprint = {https://aapm.onlinelibrary.wiley.com/doi/pdf/10.1118/1.3013703},
year = {2009}
}

@article{Frelin_dosemap_2008,
author = {Frelin, A-M. and Fontbonne, J-M. and Ban, G. and Colin, J. and Labalme, M. and Batalla, A. and Vela, A. and Boher, P. and Braud, M. and Leroux, T.},
title = {The DosiMap, a new 2D scintillating dosimeter for IMRT quality assurance: Characterization of two Čerenkov discrimination methods},
journal = {Medical Physics},
volume = {35},
number = {5},
pages = {1651-1662},
doi = {https://doi.org/10.1118/1.2897966},
url = {https://aapm.onlinelibrary.wiley.com/doi/abs/10.1118/1.2897966},
eprint = {https://aapm.onlinelibrary.wiley.com/doi/pdf/10.1118/1.2897966},
year = {2008}
}

@INPROCEEDINGS{Malvar_2004,
  author={Malvar, H.S. and Li-wei He and Cutler, R.},
  booktitle={2004 IEEE International Conference on Acoustics, Speech, and Signal Processing}, 
  title={High-quality linear interpolation for demosaicing of Bayer-patterned color images}, 
  year={2004},
  volume={3},
  pages={iii-485},
  doi={10.1109/ICASSP.2004.1326587}}

@INPROCEEDINGS{Menon_2007,
  author={Menon, Daniele and Calvagno, Giancarlo},
  booktitle={2007 IEEE International Conference on Image Processing}, 
  title={Demosaicing Based Onwavelet Analysis of the Luminance Component}, 
  year={2007},
  volume={2},
  number={},
  pages={II - 181-II - 184},
  doi={10.1109/ICIP.2007.4379122}}

@article{cloutier_deformable_2021_1,
	title = {Deformable scintillation dosimeter I: challenges and implementation using computer vision techniques},
	volume = {66},
	issn = {0031-9155, 1361-6560},
	url = {https://iopscience.iop.org/article/10.1088/1361-6560/ac1ca1},
	doi = {10.1088/1361-6560/ac1ca1},
	shorttitle = {Deformable scintillation dosimeter I},
	number = {17},
	journal = {Physics in Medicine \& Biology},
	shortjournal = {Phys. Med. Biol.},
	author = {Cloutier, E and Archambault, L and Beaulieu, L},
	urldate = {2021-08-27},
	date = {2021-09-07},
	year = 2021,
	langid = {english}
}

@article{cloutier_deformable_2021_2,
	title = {Deformable scintillation dosimeter: {II}. Real-time simultaneous measurements of dose and tracking of deformation vector fields},
	volume = {66},
	issn = {0031-9155, 1361-6560},
	url = {https://iopscience.iop.org/article/10.1088/1361-6560/ac1ca2},
	doi = {10.1088/1361-6560/ac1ca2},
	shorttitle = {Deformable scintillation dosimeter},
	pages = {175017},
	number = {17},
	journal = {Physics in Medicine \& Biology},
	shortjournal = {Phys. Med. Biol.},
	author = {Cloutier, E and Beaulieu, L and Archambault, L},
	urldate = {2021-08-27},
	date = {2021-09-07},
	year = 2021,
	langid = {english}
}

@article{therriault-proulx_method_2015,
	title = {A method to correct for temperature dependence and measure simultaneously dose and temperature using a plastic scintillation detector},
	volume = {60},
	issn = {0031-9155, 1361-6560},
	url = {https://iopscience.iop.org/article/10.1088/0031-9155/60/20/7927},
	doi = {10.1088/0031-9155/60/20/7927},
	pages = {7927--7939},
	number = {20},
	journal = {Physics in Medicine and Biology},
	shortjournal = {Phys. Med. Biol.},
	author = {Therriault-Proulx, Francois and Wootton, Landon and Beddar, Sam},
	urldate = {2021-08-26},
	date = {2015-10-21},
	langid = {english},
	year = 2015
}

@article{beddar_water-equivalent_1992,
	title        = {Water-equivalent plastic scintillation detectors for high-energy beam dosimetry: {II}. Properties and measurements},
	shorttitle   = {Water-equivalent plastic scintillation detectors for high-energy beam dosimetry},
	author       = {Beddar, A. S. and Mackie, T. R. and Attix, F. H.},
	volume       = 37,
	number       = 10,
	pages        = {1901--1913},
	issn         = {0031-9155},
	journal = {Physics in Medicine and Biology},
	shortjournal = {Phys Med Biol},
	year = 1992,
	date         = {1992-10},
	pmid         = 1438555
}

@article{beddar_water-equivalent_1992-1,
	title        = {Water-equivalent plastic scintillation detectors for high-energy beam dosimetry: I. Physical characteristics and theoretical consideration},
	shorttitle   = {Water-equivalent plastic scintillation detectors for high-energy beam dosimetry},
	author       = {Beddar, A. S. and Mackie, T. R. and Attix, F. H.},
	volume       = 37,
	number       = 10,
	pages        = {1883--1900},
	issn         = {0031-9155},
	journal = {Physics in Medicine and Biology},
	shortjournal = {Phys Med Biol},
	year = 1992,
	date         = {1992-10},
	pmid         = 1438554
}

@article{beaulieu_review_2016,
	title        = {Review of plastic and liquid scintillation dosimetry for photon, electron, and proton therapy},
	author       = {Beaulieu, Luc and Beddar, Sam},
	volume       = 61,
	number       = 20,
	pages        = {R305--R343},
	doi          = {10.1088/0031-9155/61/20/R305},
	issn         = {0031-9155, 1361-6560},
	url          = {http://stacks.iop.org/0031-9155/61/i=20/a=R305?key=crossref.88afc16233fc2bff5a63aa33e763bd8d},
	urldate      = {2016-10-07},
	journal = {Physics in Medicine and Biology},
	year = 2016, 
	date         = {2016-10-21}
}

@article{goulet_novel_2014,
	title        = {Novel, full 3D scintillation dosimetry using a static plenoptic camera},
	author       = {Goulet, Mathieu and Rilling, Madison and Gingras, Luc and Beddar, Sam and Beaulieu, Luc and Archambault, Louis},
	volume       = 41,
	number       = 8,
	pages        = {082101},
	doi          = {10.1118/1.4884036},
	issn         = {0094-2405},
	url          = {http://scitation.aip.org.acces.bibl.ulaval.ca/content/aapm/journal/medphys/41/8/10.1118/1.4884036},
	urldate      = {2016-05-31},
	journal = {Medical Physics},
	year         = 2014,
	month        = 08, 
	day          = 01
}

@article{kirov_three-dimensional_2005,
	title        = {The three-dimensional scintillation dosimetry method: test for a 106 Ru eye plaque applicator},
	shorttitle   = {The three-dimensional scintillation dosimetry method},
	author       = {Kirov, A. S. and Piao, J. Z. and Mathur, N. K. and Miller, T. R. and Devic, S. and Trichter, S. and Zaider, M. and Soares, C. G. and {LoSasso}, T.},
	volume       = 50,
	number       = 13,
	pages        = 3063,
	doi          = {10.1088/0031-9155/50/13/007},
	issn         = {0031-9155},
	url          = {http://stacks.iop.org/0031-9155/50/i=13/a=007},
	urldate      = {2016-06-02},
	journal = {Physics in Medicine and Biology},
	shortjournal = {Phys. Med. Biol.},
	year         = 2005,
	langid       = {english}
}

@article{rilling_tomographicbased_2020,
	title        = {Tomographic based {3D} scintillation dosimetry using a three view plenoptic imaging system},
	author       = {Rilling, Madison and Allain, Guillaume and Thibault, Simon and Archambault, Louis},
	year         = 2020,
	month        = may,
	journal      = {Medical Physics},
	pages        = {mp.14213},
	doi          = {10.1002/mp.14213},
	issn         = {0094-2405, 2473-4209},
	url          = {https://onlinelibrary.wiley.com/doi/abs/10.1002/mp.14213},
	urldate      = {2020-05-28},
	language     = {en}
}

@article{guillot_performance_2013,
	title        = {Performance assessment of a {2D} array of plastic scintillation detectors for {IMRT} quality assurance},
	author       = {Guillot, Mathieu and Gingras, Luc and Archambault, Louis and Beddar, Sam and Beaulieu, Luc},
	year         = 2013,
	month        = jul,
	journal      = {Physics in Medicine and Biology},
	volume       = 58,
	number       = 13,
	pages        = {4439--4454},
	doi          = {10.1088/0031-9155/58/13/4439},
	issn         = {0031-9155, 1361-6560},
	url          = {https://iopscience.iop.org/article/10.1088/0031-9155/58/13/4439},
	urldate      = {2020-04-24},
	language     = {en}
}

@article{boivin_systematic_2015,
	title        = {Systematic evaluation of photodetector performance for plastic scintillation dosimetry},
	author       = {Boivin, Jonathan and Beddar, Sam and Guillemette, Maxime and Beaulieu, Luc},
	volume       = 42,
	number       = 11,
	pages        = {6211--6220},
	doi          = {10.1118/1.4931979},
	issn         = {0094-2405},
	url          = {http://scitation.aip.org/content/aapm/journal/medphys/42/11/10.1118/1.4931979},
	urldate      = {2016-08-02},
	journal = {Medical Physics},
	date         = {2015-11},
	year = 2015, 
	langid       = {english}
}

@article{archambault_transient_2008,
	title        = {Transient noise characterization and filtration in {CCD} cameras exposed to stray radiation from a medical linear accelerator},
	author       = {Archambault, Louis and Briere, Tina Marie and Beddar, Sam},
	volume       = 35,
	number       = 10,
	pages        = {4342--4351},
	doi          = {10.1118/1.2975147},
	issn         = {2473-4209},
	url          = {http://onlinelibrary.wiley.com.acces.bibl.ulaval.ca/doi/10.1118/1.2975147/abstract},
	urldate      = {2017-11-29},
	journal = {Medical Physics},
	shortjournal = {Med. Phys.},
	date         = {2008-10-01},
	year = 2008,
	langid       = {english}
}

@article{Xue_2017_small_field,
author = {Xue, Jinyu and McKay, Jesse D. and Grimm, Jimm and Cheng, Chee-Wai and Berg, Ronald and Grimm, Shu-Ya Lisa and Xu, Qianyi and Subedi, Gopal and Das, Indra J.},
title = {Small field dose measurements using plastic scintillation detector in heterogeneous media},
journal = {Medical Physics},
volume = {44},
number = {7},
pages = {3815-3820},
doi = {https://doi.org/10.1002/mp.12272},
url = {https://aapm.onlinelibrary.wiley.com/doi/abs/10.1002/mp.12272},
year = {2017}
}

@article{archambault_mathematical_2012,
	title        = {A mathematical formalism for hyperspectral, multipoint plastic scintillation detectors},
	author       = {Archambault, Louis and Therriault-Proulx, François and Beddar, Sam and Beaulieu, Luc},
	volume       = 57, 
	number       = 21,
	pages        = 7133,
	doi          = {10.1088/0031-9155/57/21/7133},
	issn         = {0031-9155},
	url          = {http://stacks.iop.org/0031-9155/57/i=21/a=7133},
	urldate      = {2018-01-09},
	journal = {Physics in Medicine \& Biology},
	shortjournal = {Phys. Med. Biol.},
	year         = 2012,
	langid       = {english}
}

@article{therriault-proulx_development_2012,
	title        = {Development of a novel multi-point plastic scintillation detector with a single optical transmission line for radiation dose measurement},
	author       = {Therriault-Proulx, François and Archambault, Louis and Beaulieu, Luc and Beddar, Sam},
	volume       = 57,
	number       = 21,
	pages        = 7147,
	doi          = {10.1088/0031-9155/57/21/7147},
	issn         = {0031-9155},
	url          = {http://stacks.iop.org/0031-9155/57/i=21/a=7147},
	urldate      = {2018-02-09},
	journal = {Physics in Medicine \& Biology},
	shortjournal = {Phys. Med. Biol.},
	year         = 2012,
	langid       = {english}
}

@article{delage_characterization_2018,
	title        = {Characterization of a binary system composed of luminescent quantum dots for liquid scintillation},
	author       = {Delage, Marie-Ève and Lecavalier, Marie-Ève and LariviÃšre, Dominic and Allen, Claudine NÃ¬ and Beaulieu, Luc},
	volume       = 63,
	number       = 17,
	pages        = 175012,
	doi          = {10.1088/1361-6560/aad9c3},
	issn         = {0031-9155},
	url          = {https://doi.org/10.1088%2F1361-6560%2Faad9c3},
	urldate      = {2019-02-27},
	journal = {Physics in Medicine \& Biology},
	shortjournal = {Phys. Med. Biol.},
	date         = {2018-09},
	year = 2018,
	langid       = {english}
}

@article{lacroix_design_2009,
	title        = {A design methodology using signal-to-noise ratio for plastic scintillation detectors design and performance optimization},
	author       = {Lacroix, Frédéric and Beddar, A. Sam and Guillot, Mathieu and Beaulieu, Luc and Gingras, Luc},
	volume       = 36,
	number       = 11,
	pages        = {5214--5220},
	doi          = {10.1118/1.3231947},
	issn         = {0094-2405},
	url          = {https://aapm-onlinelibrary-wiley-com.acces.bibl.ulaval.ca/doi/full/10.1118/1.3231947},
	urldate      = {2019-05-18},
	journal = {Medical Physics},
	shortjournal = {Medical Physics},
	date         = {2009-11-01},
	year = 2009
}

@article{lacroix_2008_clinical,
author = {Lacroix, Fréderic and Archambault, Louis and Gingras, Luc and Guillot, Mathieu and Beddar, A. Sam and Beaulieu, Luc},
title = {Clinical prototype of a plastic water-equivalent scintillating fiber dosimeter array for QA applications)},
journal = {Medical Physics},
volume = {35},
number = {8},
pages = {3682-3690},
doi = {https://doi.org/10.1118/1.2953564},
url = {https://aapm.onlinelibrary.wiley.com/doi/abs/10.1118/1.2953564},
eprint = {https://aapm.onlinelibrary.wiley.com/doi/pdf/10.1118/1.2953564},
year = {2008}
}

@article{linares_rosales_optimization_2019,
	title        = {Optimization of a multipoint plastic scintillator dosimeter for high dose rate brachytherapy},
	author       = {Linares Rosales, Haydee M. and Duguay-Drouin, Patricia and Archambault, Louis and Beddar, Sam and Beaulieu, Luc},
	volume       = 46,
	number       = 5,
	pages        = {2412--2421},
	doi          = {10.1002/mp.13498},
	issn         = {0094-2405, 2473-4209},
	url          = {https://onlinelibrary.wiley.com/doi/abs/10.1002/mp.13498},
	urldate      = {2019-05-18},
	journal = {Medical Physics},
	shortjournal = {Med. Phys.},
	date         = {2019-05},
	year = 2019,
	langid       = {english}
}

@article{alexander_1_nodate,
	title        = {Scintillation Imaging as a High-Resolution, Remote, Versatile 2D Detection System for {MR}-Linac Quality Assurance},
	author       = {Alexander, Daniel A and Zhang, Rongxiao and Bruza, Petr and Pogue, Brian W},
	pages        = {3861-3869},
	year = 2020,
	journal = {Medical Physics},
	volume = 47,
	number = 9,
	langid       = {english}
}

@article{opencv_library,
	title        = {{The OpenCV Library}},
	author       = {Bradski, G.},
	year         = 2000,
	journal      = {Dr. Dobb's Journal of Software Tools}
}

@article{alexander_color_2021,
	title = {Color Cherenkov imaging of clinical radiation therapy},
	volume = {10},
	issn = {2047-7538},
	url = {https://doi.org/10.1038/s41377-021-00660-0},
	doi = {10.1038/s41377-021-00660-0},
	pages = {226},
	number = {1},
	journaltitle = {Light: Science \& Applications},
	shortjournal = {Light: Science \& Applications},
	author = {Alexander, Daniel A. and Nomezine, Anthony and Jarvis, Lesley A. and Gladstone, David J. and Pogue, Brian W. and Bruza, Petr},
	date = {2021-11-04}
}

@article{Jean_2021,
	doi = {10.1088/1361-6560/abf1bd},
	url = {https://doi.org/10.1088/1361-6560/abf1bd},
	year = 2021,
	month = {04},
	publisher = {{IOP} Publishing},
	volume = {66},
	number = {8},
	pages = {085009},
	author = {Emilie Jean and Fran{\c{c}}ois Therriault-Proulx and Luc Beaulieu},
	title = {Comparative optic and dosimetric characterization of the {HYPERSCINT} scintillation dosimetry research platform for multipoint applications},
	journal = {Physics in Medicine {\&} Biology},
}

@article{guillot_spectral_2011,
	title        = {Spectral method for the correction of the Cerenkov light effect in plastic scintillation detectors: A comparison study of calibration procedures and validation in Cerenkov light-dominated situations: Cerenkov light correction in {PSDs}},
	shorttitle   = {Spectral method for the correction of the Cerenkov light effect in plastic scintillation detectors},
	author       = {Guillot, Mathieu and Gingras, Luc and Archambault, Louis and Beddar, Sam and Beaulieu, Luc},
	volume       = 38,
	number       = 4,
	pages        = {2140--2150},
	doi          = {10.1118/1.3562896},
	issn         = {00942405},
	url          = {http://doi.wiley.com/10.1118/1.3562896},
	urldate      = {2020-09-04},
	journal = {Medical Physics},
	shortjournal = {Med. Phys.},
	date         = {2011-03-24},
	year = 2011,
	langid       = {english}
}
\end{small}

\end{document}